\documentclass[letter,scriptaddress,twocolumn, prb,showkeys]{revtex4}
	\usepackage{amsmath, amssymb} 
	\usepackage{makeidx}
	\usepackage{amsfonts}
	\usepackage{dsfont}
	\usepackage{subfigure}
	\usepackage{epsfig}
	\usepackage[colorlinks,hyperindex]{hyperref}
	\hypersetup
	{
		colorlinks,%
		citecolor=black,%
		linkcolor=black,%
		urlcolor=black,%
	}

\makeindex

\begin{document}

\title{Quantum Electrodynamics in $d=3$ from the $\epsilon$-expansion}

\author{Lorenzo Di Pietro}
\email{lorenzo.dipietro@weizmann.ac.il}
\author{Zohar Komargodski}
\email{zohar.komargodski@weizmann.ac.il}
\author{Itamar Shamir}
\email{itamar.shamir@weizmann.ac.il}
\author{Emmanuel Stamou}
\email{emmanuel.stamou@weizmann.ac.il}
\affiliation{Department of Particle Physics and Astrophysics \\ Weizmann Institute of Science, Rehovot 7610001, Israel}

\date{August 26, 2015}


\begin{abstract}

We study Quantum Electrodynamics in $d=3$ (QED$_3$) coupled to $N_f$ flavors of fermions. The theory flows to an IR fixed point for $N_f$ larger than some critical number $N_f^c$. For $N_f\leq N_f^c$, chiral-symmetry breaking is believed to take place. In analogy with the Wilson-Fisher description of the critical $O(N)$ models in $d=3$, we make use of the existence of a fixed point in $d= 4 - 2\epsilon$ to study the three-dimensional conformal theory. We compute in perturbation theory the IR dimensions of fermion bilinear and quadrilinear operators. For small $N_f$, a quadrilinear operator can become relevant in the IR and destabilize the fixed point. Therefore, the $\epsilon$-expansion can be used to estimate $N_f^c$. An interesting novelty compared to the $O(N)$ models is that the theory in $d=3$ has an enhanced symmetry due to the structure of 3d spinors.  We identify the operators in $d=4-2\epsilon$ that correspond to the additional conserved currents at $d=3$ and compute their infrared dimensions. 

\end{abstract}

\keywords{strongly correlated electrons, chiral-symmetry breaking, quantum electrodynamics, $\epsilon$-expansion, RG flow}

\maketitle


\section{Introduction}

We consider an $\mathbb{R}$ gauge theory in $d=2+1$ dimensions, coupled to $2 N_f$ complex two-component massless fermions of unit charge, $\psi^i$ ($i = 1, \ldots, 2N_f$). This theory has an $SU(2N_f)$ global symmetry.\footnote{Since we are discussing the $\mathbb{R}$ gauge theory, the topological symmetry current $j=\star F$, where $F$ is the field strength two-form, does not have any local operators charged under it and can thus be ignored for the time being.} When $N_f$ is sufficiently large, the theory flows to a stable interacting fixed point with an $SU(2N_f)$ global symmetry.
(It is stable in the sense that there are no relevant operators preserving all the symmetries.)
However, the IR behavior is different if the number of fermions is smaller than a critical value $N_f \leq N_f^c$, leading to spontaneous symmetry breaking according to the pattern~\cite{Vafa:1984xh}
\begin{equation}\label{pattern}
SU(2N_f) \to SU(N_f) \times SU(N_f) \times U(1) ~.
\end{equation}
This symmetry breaking pattern can be triggered by the condensation of the parity-even operator \begin{equation}\label{chiralcond}\sum_{a=1}^{N_f}\left(\bar \psi_{a} \psi^{a} - \bar \psi_{a + N_f} \psi^{a + N_f}\right)~.\end{equation} 
Various estimates of the critical number $N_f^c$ exist in the literature.

In condensed matter physics this theory has been advocated as an effective description of various strongly correlated materials. 
QED$_3$ can arise as the continuum limit of spin systems with various values of $N_f$, e.g. $N_f=2,4$.\cite{Marston:1989zz,Ran et al.(2007),Hermele et al.(2008)} 
The theory with $N_f = 2$ has also applications in high-temperature superconductivity.\cite{Rantner & Wen(2001), RantnerWen(2002),Hermele et al.(2005)} 

A method that has been employed to study QED$_3$ is the large-$N_f$ expansion.\cite{Pisarski:1984dj, Appelquist:1986fd, Appelquist:1988sr, Appelquist:2004ib} At large $N_f$, the theory simplifies  and a systematic expansion in $1/N_f$ can be carried out.\footnote{The theory simplifies at $N_f=\infty$ for all $d=4-2\epsilon$. This was considered in \,\cite{Gracey:1993sn, Gracey:1993iu}.} For an alternative to large $N_f$ that uses the functional renormalization group approach see \,\cite{Braun:2014wja}.

Here, we study QED$_3$ using the $\epsilon$-expansion. 
Clearly, since the theory is IR free in $d=4$ and since the gauge coupling 
has positive mass dimension for $d<4$, there is an IR fixed point at 
$d=4-2\epsilon$ with $\epsilon>0$. 
The fixed point is generated analogously to the Wilson-Fisher fixed point.\cite{Wilson:1971dc} Experience with the Wilson-Fisher fixed point of $O(N)$ models, on which both
$\epsilon$ and large-$N$ expansions have been applied,\footnote{For instance, the Wilson-Fisher prediction for the dimension of the energy 
operator is 
$\Delta(\phi^2)=d-2+\frac{2N+4}{N+8}\epsilon+\cdots$. 
We can contrast this with the large-$N$ prediction
$\Delta(\phi^2)=2-\frac{1.08}{N}+\cdots$. 
The $\epsilon$-expansion accounts for the known dimension in $d=3$, $N=1$, $\Delta(\phi^2)=1.41\dots$ to within 5\%, while the large-$N$ prediction is not as good.} 
suggests that in QED the $\epsilon$-expansion may be more effective in describing small-$N_f$ physics.

In the development of the $\epsilon$-expansion for QED one encounters some new technical difficulties that do not arise for $O(N)$ models. Perhaps one reason that (to our knowledge) it has not been considered before is that spinor representations of the Poincar\'e group do not behave very simply as a function of the number of dimensions (unlike tensor representations).\footnote{For previous applications of the $\epsilon$-expansion to fermionic systems see \,\cite{Ponte:2012ru, Dalidovich:2013qta, Patel}.}
Hence, it may not be obvious how to analytically  continue to $d$. 
However, there appears to be no fundamental obstruction to studying QED with $d\leq 4$. We will keep the spinor structure that exists in $d=4$ also in lower dimension. In lower integer dimension, the representation is reducible and can be interpreted in terms of the existing spinor structures in $d=3$ and $d=2$. 

The fact that spinor representations are smaller in $d=3$ than in $d=4$ enhances the symmetry of the theory. The theory in $d=3$ enjoys an $SU(2N_f)$ global symmetry, while the theory in $d=4$ (around which we expand) only an $SU(N_f) \times SU(N_f)$ symmetry. We find that in $d=4-2\epsilon$ certain antisymmetric tensor operators, bilinear in the fermions, are naturally interpreted as continuations of the enhanced currents of the three-dimensional theory. This suggests that the $\epsilon$-expansion provides the necessary elements to correctly describe the theory in $d=3$. 

Here, we only perform leading-order computations in the $\epsilon$-expansion of QED. Going to higher orders will be necessary to acquire more confidence about the accuracy of the method, and to estimate the uncertainties.\cite{DiPietro:2015xx} We consider bilinear and quadrilinear operators in the fermions. We shall see that a certain quadrilinear operator invariant under $SU(2N_f)$ and parity can become relevant in the IR for low values of $N_f$, and may destabilize the fixed point. At leading order in $\epsilon$, evaluating the dimension naively at $\epsilon=1/2$ without any resummation leads to $N_f^c = 2$. This is consistent both with the $F$-theorem~\cite{Grover:2012sp} and with lattice data.\cite{Hands:2002dv, Hands:2004bh, Strouthos:2008kc, Raviv:2014xna} (A different estimate that uses input from the $d=2$ flavored Schwinger model gives $N_f^c = 4$.\footnote{We thank S.~Giombi, I.~Klebanov, and G.~Tarnopolsky for conversations on how to use input from $d=2$.}) We also estimate the dimensions of these bilinear and quadrilinear operators at the fixed point for $N_f>N_f^c$. 


\subsection{Compact vs Non-Compact Gauge Group}

In this note we do not study QED$_3$ with compact gauge group $U(1)$,\cite{Aharony:2015xx} 
but we would like to make several comments nevertheless.\footnote{We thank N.~Seiberg and S.~Pufu for discussions on this matter.} The compact theory has monopole operators that sit in various representations of $SU(2N_f)\times U(1)_T$, where $U(1)_T$ is associated to the topologically conserved current $j_\mu=\epsilon_{\mu\nu\rho} F^{\nu\rho}$. Monopole operators can condense (i.e. have an expectation value in the vacuum) and they can proliferate (meaning that we can add them to the action). Because monopole operators are charged under the global symmetry group, the question of monopole proliferation is sensitive to the symmetries preserved by the regulator (be it the lattice or some other UV theory that flows to QED$_3$). Therefore, in QED$_3$  monopole proliferation has presumably nothing to do with chiral-symmetry breaking, which is an intrinsic phenomenon, given that it also happens in the non-compact theory, which has no monopoles. In fact, if monopole operators proliferate, the Nambu-Goldstone bosons of~\eqref{pattern} would generically become massive because the $SU(2N_f)$ symmetry would be broken explicitly. This mechanism of gapping the theory is analogous to the mechanism for confinement in Polyakov's model.\cite{Polyakov} On the other hand, we do expect monopole operators to condense. The lightest monopole operator sits in the rank-$N_f$ antisymmetric representation, and its condensation would be precisely consistent with the symmetry breaking pattern~\eqref{pattern}.\footnote{For this, the vacuum-expectation value of the monopole operator needs to be aligned with the vacuum-expectation value of the fermion bilinear~\eqref{chiralcond}. Otherwise, the global symmetry would be broken further. }


\section{The $\epsilon$-expansion}

\subsection{Generalities}

To illustrate the procedure of the $\epsilon$-expansion, consider the two-point function of an operator $\mathcal{O}$ in $d=4$, expanded in perturbation theory in a classically marginal coupling $g$
\begin{equation}\label{exp4}
\langle \mathcal{O}(p) \mathcal{O}(-p)\rangle = p^{2\Delta - 4} 
\!\!
\sum_{0 \leq m \leq n, n = 0}^\infty \!\!c_{nm}\, g^n \left( \log \frac{\Lambda^2}{p^2} \right)^m\!\!\!, 
\end{equation}
where $\Delta$ is the dimension of $\mathcal{O}$ in $d=4$ at $g=0$, and $\Lambda$ is a UV cutoff. Introducing the renormalized operator $\mathcal{O}^{ren} = Z \mathcal{O}$, we can cancel the $\Lambda$-dependence of the correlator by allowing the coupling $g$ and the normalization $Z$ to evolve according to
$\frac{d g}{d \log \Lambda} \equiv \beta(g)$, $\frac{d \log Z}{d \log \Lambda} \equiv \gamma(g)$,
such that the Callan-Symanzik equation holds
\begin{equation}\label{CS}
\left(\frac{\partial}{\partial \log \Lambda} + \beta(g) \frac{\partial}{\partial g} + 2\gamma(g) \right) \langle \mathcal{O}(p) \mathcal{O}(-p)\rangle = 0~.
\end{equation}

The terms in \eqref{exp4} with coefficients $c_{nn}$, $n\geq 1$, are the leading logs. It follows from~\eqref{CS} that they are all fixed  in terms of the coefficients $\beta_1$ and $\gamma_1$ in the leading order expansion of $\beta$ and $\gamma$
\begin{equation}
\beta(g) = \beta_1 g^2 + O(g^3)\,,\quad \gamma(g) = \gamma_1 g + O(g^2)~.
\end{equation}
One can then resum the leading logs to obtain 
\begin{equation}\label{CSprefactor}
\langle{\cal O}(p){\cal O}(-p)\rangle \simeq p^{2\Delta -4} \left(1 + \frac 12 \beta_1 g \log\frac{\Lambda^2}{p^2}	\right)^{- \frac{2 \gamma_1}{\beta_1}} ~. 
\end{equation}

For $d= 4 -2 \epsilon$ we assume that $g$ acquires a positive mass dimension $c\epsilon$ (where $c$ is some positive number). The analogue perturbative expansion of the two-point function in $d = 4-2\epsilon$ is
 \begin{equation}\label{expd}
\langle \mathcal{O}(p)\mathcal{O}(-p)\rangle = p^{2\Delta - d} \sum_{n = 0}^\infty c_n\, \left(\frac{g}{p^{c\epsilon}}\right)^n ~. 
\end{equation}
Requiring that \eqref{expd} approaches \eqref{exp4} in the limit $\epsilon \to 0$, we find the matching condition
\begin{equation}
c_n = \sum_{m = 0}^n c_{nm} \, m! \left(\frac{2}{c \epsilon}\right)^m + O(\epsilon)~.
\end{equation}
The leading contribution to the two-point function  \eqref{expd} in the limit $\epsilon \ll 1$ comes from the terms containing $c_{nn}$, which we can thus resum similarly to \eqref{CSprefactor}
\begin{equation}\label{CSprefd}
\langle{\cal O}(p){\cal O}(-p)\rangle \simeq
p^{2\Delta-d} \left(1 + \frac{\beta_1}{c \epsilon}  \frac{g}{p^{c\epsilon}}	\right)^{- \frac{2 \gamma_1}{\beta_1}} 
\!\!\!\!\!\underset{\overset{p \to 0}{}}{\approx} p^{2\Delta-d}p^{2 \gamma_1\frac{c\epsilon}{\beta_1}}.
\end{equation}
In the IR limit $p \to 0$ a new scaling law emerges. The contribution to the IR dimension of the operator at first order in $\epsilon$ is thus
\begin{equation}\label{DeltaIR}
\Delta_{\rm IR}  = \Delta + \gamma_1 \frac{c\epsilon}{\beta_1} + O(\epsilon^2)\,.
\end{equation}
The crossover to the IR scaling in \eqref{CSprefd} happens when
\begin{equation}
1 \ll \frac{\beta_1}{c \epsilon}  \frac{g}{p^{c\epsilon}} \quad\Rightarrow \quad p \ll\left(\frac{\beta_1}{c \epsilon}\right)^{\frac{1}{c\epsilon}} g^{\frac{1}{c\epsilon}}~.
\end{equation}
We see here the physical consequence of introducing the small parameter $\epsilon$: the crossover towards the 
IR happens at a scale that is enhanced by the parametrically large factor $\left(\beta_1 / c \epsilon\right)^{\frac{1}{c\epsilon}}$ 
with respect to the naive scale $g^{\frac{1}{c\epsilon}}$. As a result, the IR fixed point is parametrically close to the one in the UV.

Indeed, the IR fixed point corresponds to a zero of the $\beta$-function for the dimensionless combination $\hat{g} = g\Lambda^{-c\epsilon}$
\begin{align}
\frac{d \hat{g}}{d \log \Lambda} \equiv \beta(\hat{g}) = -c \epsilon \, \hat{g} + \beta_1 \hat{g}^2 + O(\hat{g}^3) \nonumber\\
\beta(\hat{g}_*)= 0 \, \Rightarrow \, \hat{g}_* = \frac{c \epsilon}{ \beta_1} + O(\epsilon^2)~.
\end{align}
Comparison with \eqref{DeltaIR} shows explicitly that, at leading order in $\epsilon$, 
the difference $\Delta_{\rm IR} - \Delta$ is the anomalous dimension $\gamma$ evaluated at the fixed point.\cite{Brezin:1973jc, Brezin:1974eb} 
Extrapolating the results to $\epsilon = \frac 12$, we obtain an estimate for the observables of the IR theory in three dimensions.


\subsection{Wilson-Fisher Fixed Point in QED}

The Lagrangian for QED in $d = 4$ is
\begin{equation}\label{Lagd}
\mathcal{L} = -\frac{1}{4 e^2} F^{\mu\nu} F_{\mu\nu} + i\sum_{a = 1}^{N_f} \bar{\Psi}_a \gamma^\mu D_\mu \Psi^a~.
\end{equation}
We use the usual four-dimensional Dirac notation for the spinors. 
Their decomposition in terms of two-component fermions is
\begin{align}
\Psi^a & = \left(\begin{array}{c} \psi^{a} \\ i\sigma_2\psi^{a+N_f} \end{array}\right)~,\quad a = 1,\dots, N_f~.
\end{align}

In dimension $d$ we take the Clifford algebra to be $\{\gamma^\mu,\gamma^\nu\} = 2 \eta^{\mu\nu} \mathds{1}$, with $\eta^{\mu\nu}\eta_{\mu\nu} = d$.
To leading nontrivial order, the beta function in $d=4-2\epsilon$ is given by (here $\hat{e} = e \Lambda^{-\epsilon}$)
\begin{equation}\label{beta}
\beta(\hat{e}) = - \epsilon \hat{e} + \frac{N_f}{12 \pi^2} \hat{e}^3 + O(\hat{e}^5)~. 
\end{equation}
The value of the coupling at the Wilson-Fisher fixed point is $\hat{e}^2_* = 12 \pi^2  \epsilon/N_f$.
The theory is therefore weakly coupled when we are close to $d=4$ or when the number of flavors is large.

A comment on $\gamma_5$ is in order. A consistent definition of $\gamma_5$ in non-integer dimension is due to 't Hooft and Veltman.\cite{'tHooft:1972fi,Breitenlohner:1977hr,Akyeampong:1973vj}
According to this prescription, $\gamma_5$ anticommutes only with the $\gamma^\mu$'s of the four-dimensional subspace, 
and commutes with all others. This implies an explicit breaking of axial symmetries in $d=4-2\epsilon$, and 
reproduces the chiral anomaly for the singlet axial current $\sum_a\bar{\Psi}_a\gamma^\mu \gamma_5 \Psi^a$ in the limit 
$\epsilon \to 0$.\cite{Collins, Larin:1993tq} For the leading-order calculations that we present here, 
this prescription is in practice equivalent to a naive continuation of $\gamma_5$ as totally anticommuting. 
However, the difference from the naive continuation becomes relevant at higher orders. 

QED in $d=4$ has an $SU(N_f)\times SU(N_f)$ global symmetry with associated conserved currents
\begin{equation}
\label{}
\begin{split}
(J_\mu)_a^{~b} &= \bar{\Psi}_a \gamma_\mu \Psi^b - \frac{1}{N_f}\delta_a^{~b}\sum_c \bar{\Psi}_c \gamma_\mu \Psi^c ~,\\
(J^5_\mu)_a^{~b} &=\bar{\Psi}_a \gamma_\mu \gamma_5 \Psi^b  - \frac{1}{N_f}\delta_a^{~b}\sum_c \bar{\Psi}_c \gamma_\mu\gamma_5 \Psi^c~.
\end{split}
\end{equation}
Their anomalous dimension at one-loop vanishes and, therefore, at leading order the IR dimension is the same as the classical one, i.e. $d-1$. This is the correct scaling dimension for conserved currents. For the vector current this argument is valid at all orders in perturbation theory, because they are conserved for any $d$. On the other hand, the axial currents $J_\mu^5$ are explicitly broken for non-integer $d$,\cite{Collins} and this can affect the IR dimension at higher orders. Nevertheless, we do expect them to be conserved in $d=3$, because the non-conservation is given by an operator that vanishes both in $d=4$ and $d=3$. 

So far, we have argued that the $\epsilon$-expansion predicts the existence of currents associated to the global symmetry $SU(N_f) \times SU(N_f)$ 
in the IR CFT for $d=3$. 
However, QED$_3$ has an enhanced $SU(2N_f)$ symmetry. For $N_f \geq N_f^c$, the full $SU(2N_f)$ is realized linearly at the IR fixed point. 
This entails the existence of $2N_f^2 + 1$ additional conserved operators of spin 1 with protected 
dimension $\Delta = 2$. 
It is natural to ask whether these operators are visible also in the theory continued to non-integer dimension.

One of the additional currents is the singlet axial current $J_\mu^s = \sum_a \bar{\Psi}_a \gamma_\mu \gamma_5 \Psi^a$. 
Indeed, the continuation of the anomaly operator $F\wedge F$ vanishes for $d=3$.
As for the remaining $2N_f^2$ currents, we note that in $d = 4-2\epsilon$ we can define the following antisymmetric tensor operators
\begin{equation} (K_{\mu\nu})_a^{~b} = \bar{\Psi}_a \gamma_{\mu\nu} \Psi^b~,~\bar{\Psi}_a \gamma_{\mu\nu} \gamma_5 \Psi^b~.
\end{equation} 
They carry the correct flavor and Lorentz quantum numbers to be identified with the additional currents, 
because in $d=3$ we can use the totally antisymmetric tensor $\epsilon_{\mu\nu\rho}$ and dualize them to spin 1 operators. 

We are led to the expectation that the IR dimension of $J_\mu^s$ and $K_{\mu\nu}$ should evaluate to 2 for $\epsilon = 1/2$. 
The one-loop computation gives 
\begin{equation}
\label{dimtwoform}
\begin{split}
\Delta_{\rm IR}(J^s) & = 3 - 2 \epsilon + O(\epsilon^2)\,,\\
\Delta_{\rm IR}(K) & = 3 - 2 \epsilon + \frac{3 \epsilon}{2N_f} + O(\epsilon^2)\,.
\end{split}
\end{equation}
The anomalous dimension of $J^s$ only starts at two-loop order.\cite{Kodaira:1979pa, Larin:1993tq} 
As we will show in the next section, we can estimate that the IR critical point exists only for $N_f \geq 3$. Plugging $N_f = 3$ and $\epsilon = 1/2$ in \eqref{dimtwoform} we find $\Delta^{\text{1-loop}}_{\rm IR}(K) = 2.25$, which agrees with the expectation within a 10\% margin. The precision improves for larger values of $N_f$. We view this as a hint that the continuation to non-integer dimensions correctly captures the properties of the 3d CFT that we ultimately want to study. A preliminary check of higher orders in $\epsilon$ shows that the agreement {\it improves}. This will be discussed elsewhere.

Note that we can also study the anomalous dimension of the operator $F_{\mu\nu}$.  
The Bianchi identity is obeyed for all $d$, and one can verify that  $\Delta_{\rm IR}(F) = 2$ holds to all orders in $\epsilon$.

\subsection{Quadrilinear and Bilinear Operators}

In the three-dimensional theory there are two parity-even quadrilinear scalar operators that are invariant under the full $SU(2N_f)$ 
\begin{equation}
\label{O12}
\mathcal{O}_1 = \bigl(\sum_i \bar \psi_{i} \sigma^\mu \psi^{i}
\bigl)^2
\quad\text{and}\quad
\mathcal{O}_2 = \bigl(\sum_i \bar \psi_{i} \psi^{i} \bigl)^2 \,. 
\end{equation}
The operator $\mathcal{O}_1$ can be easily continued to $d = 4-2\epsilon$. 
In Dirac notation we can rewrite it as $\mathcal{O}_1 =(\sum_a \bar{\Psi}_a\gamma_\mu \Psi^a)^2$. 
To continue the operator $\mathcal{O}_2$, we use the fact that in $d=3$ the antisymmetrization of three $\gamma$ matrices is proportional to the identity.
Therefore, we can rewrite it as $\mathcal{O}_2 = 6 (\sum_a\bar{\Psi}_a\gamma_{[\mu}\gamma_\nu\gamma_{\rho]} \Psi^a)^2$, 
which is a well-defined expression also for $d=4-2\epsilon$. Note that in $d=4$ this operator can be identified with the square of the axial current $(\sum_a\bar{\Psi}_a\gamma_\mu \gamma_5 \Psi^a)^2$.

\begin{figure}[htbp]
 	\begin{center}
	\includegraphics[]{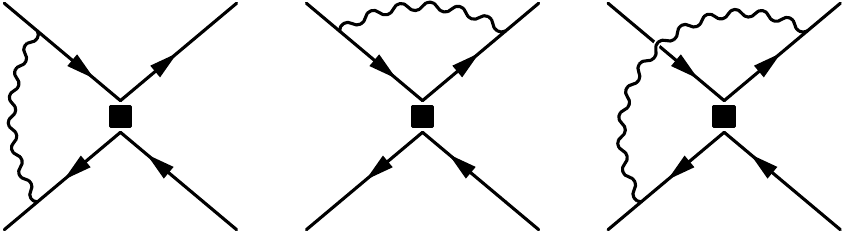} 	
	\caption{Diagrams giving the mixing matrix of the quadrilinear operators at one-loop.}
	\label{fig:quadrilinear}
	\end{center}
\end{figure}

\begin{figure}[htbp]
 	\begin{center}
	\includegraphics[]{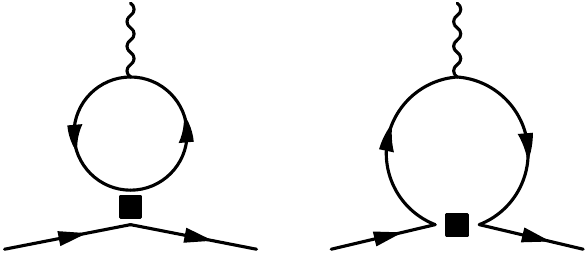}
	\caption{Diagrams giving the mixing matrix of $\mathcal{O}_1$ and $\mathcal{O}_2$ into $\mathcal{O}_{\rm EOM}$.}
	\label{fig:EOM}
	\end{center}
\end{figure}

To obtain their IR dimension we compute the mixing between these two operators. 
Typical diagrams at one-loop are shown in Fig \,\ref{fig:quadrilinear}.
To obtain the correct mixing at one-loop, it is necessary to also take into account 
the one-loop mixing with the operator $\mathcal{O}_{\rm EOM} = (\sum_a \bar{\Psi}_a\gamma^\mu \Psi^a)(\tfrac 1 e \partial^\nu F_{\mu\nu} - \sum_b \bar{\Psi}_b \gamma_\mu \Psi^b)$ that vanishes on the equations of motion. This mixing is induced by the diagrams in Fig \,\ref{fig:EOM}.

In the basis $\{ \mathcal{O}_1, \mathcal{O}_2\}$, the matrix of anomalous dimensions reads~\cite{Beneke:1995qq}
\begin{equation}\label{matranodim}
\gamma_\mathcal{O}(\hat{e}) = \frac{\hat{e}^2}{16 \pi^2} \left(\begin{array}{cc} \frac{8}{3}(2N_f + 1) & 12 \\ \frac{44}{3} & 0 \end{array}\right) + O(\hat{e}^4)~.
\end{equation}
Its eigenvalues are $\frac{\hat{e}^2}{12 \pi^2}(2N_f +1 \pm 2\sqrt{N_f^2 + N_f + 25})$.\footnote{Note that the matrix $\gamma_\mathcal{O}$ becomes symmetric in the basis of operators $\{ \tilde{\mathcal{O}}_1 = \tfrac{1}{\sqrt{N_f}} (-\mathcal{O}_1 + \mathcal{O}_2),~~\tilde{\mathcal{O}}_2 = \tfrac{1}{\sqrt{N_f +1}}(\mathcal{O}_1 + \mathcal{O}_2)\}$, whose matrix of tree-level two-point functions is proportional to the identity.}

By evaluating them for the fixed-point value $\hat{e}^2_*=12\pi^2\epsilon/N_f$, 
we find that the operator corresponding to the negative eigenvalue 
becomes relevant when $N_f \leq \frac{9\epsilon}{2} + O(\epsilon^2)$. 
Thus, for $\epsilon = \frac 12$ the operator is relevant in the IR when $N_f = 1,2$, 
while it remains irrelevant for all integer $N_f > 2$. 
Because the quadrilinear is neutral under all the global symmetries, 
when it becomes relevant it may be generated and trigger a flow to a new IR 
phase (e.g. a Goldstone phase). From this we obtain the estimate $N_f^c \leq 2$. 

The same mechanism for the onset of chiral-symmetry breaking has been studied using the large-$N_f$ expansion.\cite{Xu(2008)} The large-$N_f$ approximation of the anomalous dimensions is such that $\mathcal{O}_1$ and $\mathcal{O}_2$ are always irrelevant at the IR fixed point for $N_f \geq 1$. (See however the renormalization group study at large $N_f$ in \,\cite{Kaveh:2004qa}.)

Let us make a few comments on $d=2$. There, the quadrilinear operator is marginal already at the tree-level. Therefore, the criterion above implies that in $d=2$ the anomalous dimension must evaluate to 0 at $N_f^c$. This is satisfied for $N_f^c = \infty$. This value is consistent with the IR behavior of QED in $d=2$: for every finite $N_f$ the theory flows to an $SU(2N_f)$ Wess-Zumino-Witten (WZW) interacting CFT,\cite{Gepner:1984au, Affleck:1985wa} namely a $\sigma$-model with coset target space deformed by a WZW term (exactly as one expects for the Lagrangian of Nambu-Goldstone bosons). Even though Nambu-Goldstone bosons do not exist in $d=2$, it appears natural to interpret this theory as the continuation of the chirally broken phase to $d=2$. See also \,\cite{Witten:1983ar, Polyakov:2005ss}. Assuming that the anomalous dimension approaches 0 as $1/N_f$ for $N_f \to \infty$, the divergence of $N_f^c(d)$ for $d = 2$ is given by a simple pole. This suggests using the modified ansatz $N_f^c(d)= (d-2)^{-1} f(d)$ in the equation for $N_f^c$, which can then be solved for $f$ perturbatively in $\epsilon$. With this ansatz, the leading-order estimate becomes $N_f^c \leq 4.5$. The difference with the previous estimate are higher-order terms in $\epsilon$; it may be viewed as a measure of uncertainty. Improving on this requires computations beyond one loop.\cite{DiPietro:2015xx}

\begin{figure}[htbp]
 	\begin{center}
	\includegraphics[]{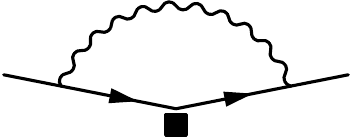}
	\caption{The diagram giving the anomalous dimension of bilinear operators at one-loop.}
		\label{fig:bilinear}
	\end{center}
\end{figure}

Further data about the fixed point can be obtained by considering bilinear scalar operators. There are two types of scalar operators in the three-dimensional theory. Operators of the first type are scalars also in $d\neq 3$, i.e.
\begin{equation}
(B_1)_a^{~b} = \bar{\Psi}_a \Psi^b~,~\bar{\Psi}_a \gamma^5\Psi^b~.
\end{equation}
Operators in this class preserve at most the diagonal $SU(N_f)$ subgroup of $SU(N_f)\times SU(N_f)$, 
and the most symmetric ones are
$\sum_a \bar{\psi}_a\psi^{a+N_f} \pm c.c.$.
The one-loop computation (see Fig\,\ref{fig:bilinear}) gives
\begin{equation}
\Delta_{\rm IR}(B_1) = 3 - 2 \epsilon - \frac{9 \epsilon}{2 N_f} + O(\epsilon^2)~.
\end{equation}

The second type of scalar operators are given by rank-three antisymmetric tensors  in $d=4-2\epsilon$
\begin{equation}
({B_2}_{\mu\nu\rho})_a^{~b} = \bar{\Psi}_a \gamma_{[\mu}\gamma_{\nu} \gamma_{\rho]}\Psi^b~,~\bar{\Psi}_a \gamma_{[\mu}\gamma_{\nu} \gamma_{\rho]}\gamma^5\Psi^b~.
\end{equation}
They give rise to scalars in $d=3$ because they can be contracted with the totally antisymmetric tensor $\epsilon_{\mu\nu\rho}$. The chiral condensate \eqref{chiralcond} and the parity-odd, $SU(2N_f)$-invariant bilinear $\sum_i \bar{\psi}_i \psi^i$ belong to this class of operators. Since their anomalous dimension vanishes at leading order in perturbation theory, their IR dimension to first order in $\epsilon$ is captured by just the classical contribution
\begin{equation}\label{andimB2}
\Delta_{\rm IR}(B_2) = 3 - 2 \epsilon + O(\epsilon^2)~.
\end{equation}
The anomalous dimension starts being non-zero at two-loop order.\cite{Larin:1993tq,Gracey:2000am} This implies that higher orders in $\epsilon$ in \eqref{andimB2} will be non-zero. Nevertheless, the $\epsilon$-expansion suggests that the IR dimension of these scalar operators is perhaps close to $\Delta = 2$.

\section{Future Directions}

In this paper we initiated a study of the critical point of QED$_3$ based on the $\epsilon$-expansion. Our results were based on leading-order computations. It would be very interesting to sharpen the theoretical predictions by higher-order computations.\footnote{
At higher orders it becomes necessary to take into account ``evanescent operators''~\cite{Dugan:1990df} that vanish in integer dimension, 
but can mix with physical operators. We thank S.\,Rychkov for discussions on this point. Similar phenomena are familiar in the context of 
Gross-Neveu models in $d=2+\epsilon$.\cite{Bondi:1989nq}} The necessary preliminary step of computing the two-loops counterterms was done for generic gauge theories with fermions in \,\cite{Jack:1984vj}. Due to the asymptotic nature of the $\epsilon$-expansion, efficiently including higher-order terms requires the use of resummation techniques. (In this context, it would be interesting to understand if data from the flavored Schwinger model can be efficiently included.)
 
The $\epsilon$-expansion can be used to compute additional observables of the IR fixed point. For instance, correlators of the stress tensor and of conserved currents.\footnote{For the large-$N_f$ study see \,\cite{Huh:2014eea, Chowdhury:2012km}.} Another interesting datum of the 3d theory is the universal coefficient $F$ of the partition function on the three-sphere, which gives also the universal part of the entanglement entropy across a circular region. For this one can utilize the techniques of \,\cite{Giombi:2014xxa, Fei:2015oha}. The calculation of $F$ in QED$_3$ via the $\epsilon$-expansion has been recently presented in \,\cite{Giombi:2015haa}. Another interesting line of investigation would be to see if the conformal bootstrap techniques shed any light on QED$_3$.\cite{Rattazzi:2008pe} 

\begin{acknowledgements}
 
We thank Ofer Aharony, Leon Balents, Jan de Boer, Holger Gies, Simone Giombi,  Igor Klebanov, Sung-Sik Lee, Yu Nakayama, Prithvi Narayan, Hugh Osborn, Silviu Pufu, Slava Rychkov, Subir Sachdev, Adam Schwimmer, Nathan Seiberg, Tarun Sharma, Philipp Strack,
and Grigory Tarnopolsky, for useful discussions. This work was supported in part by an Israel Science Foundation center for excellence grant, by the I-CORE program of the Planning and Budgeting Committee and the Israel Science Foundation (grant number 1937/12), by the ERC STG grant 335182, and by the United States-Israel Bi-national Science Foundation (BSF) under grant 2010/629. 

\end{acknowledgements}


\begin{thebibliography}{}



\bibitem{Vafa:1984xh}
C.~Vafa and E.~Witten,
``Eigenvalue Inequalities for Fermions in Gauge Theories,''
Commun.\ Math.\ Phys.\ {\bf 95} (1984) 257.

\bibitem{Marston:1989zz}
J.~B.~Marston and I.~Affleck,
``Large-N Limit of the Hubbard-Heisenberg Model,''
Phys.\ Rev.\ B {\bf 39} (1989) 11538.

\bibitem{Ran et al.(2007)} 
Y.~Ran, M.~Hermele, P.~A.~Lee and X.~G.~Wen,
``Projected-Wave-Function Study of the Spin-1/2 Heisenberg Model on the Kagome Lattice,''
Phys.\ Rev.\ Lett.\  {\bf 98} (2007) 11,  117205.

\bibitem{Hermele et al.(2008)} 
  M.~Hermele, Y.~Ran, P.~Lee and X.~G.~Wen,
  ``Properties of an algebraic spin liquid on the kagome lattice,''
  Phys.\ Rev.\ B {\bf 77}, no. 22, 224413 (2008).

\bibitem{Rantner & Wen(2001)} 
  W.~Rantner and X.~G.~Wen,
  ``Electron Spectral Function and Algebraic Spin Liquid for the Normal State of Underdoped High $T_c$ Superconductors,''
  Phys.\ Rev.\ Lett.\  {\bf 86} (2001), 3871.


\bibitem{RantnerWen(2002)} 
  W.~Rantner and X.~G.~Wen,
  ``Spin correlations in the algebraic spin liquid: Implications for high-$T_c$ superconductors,''
  Phys.\ Rev.\ B {\bf 66}, 144501 (2002).

\bibitem{Hermele et al.(2005)} 
M.~Hermele, T.~Senthil, and M.~P.~A.~Fisher, 
``Algebraic spin liquid as the mother of many competing orders,''
  Phys.\ Rev.\ B {\bf 72}, 104404 (2005).

\bibitem{Pisarski:1984dj} 
  R.~D.~Pisarski,
  ``Chiral Symmetry Breaking in Three-Dimensional Electrodynamics,''
  Phys.\ Rev.\ D {\bf 29}, 2423 (1984).

\bibitem{Appelquist:1986fd} 
  T.~W.~Appelquist, M.~J.~Bowick, D.~Karabali and L.~C.~R.~Wijewardhana,
  ``Spontaneous Chiral Symmetry Breaking in Three-Dimensional QED,''
  Phys.\ Rev.\ D {\bf 33}, 3704 (1986).
  
\bibitem{Appelquist:1988sr}
  T.~Appelquist, D.~Nash and L.~C.~R.~Wijewardhana,
  ``Critical Behavior in (2+1)-Dimensional QED,''
  Phys.\ Rev.\ Lett.\  {\bf 60} (1988) 2575.

\bibitem{Appelquist:2004ib}
  T.~Appelquist and L.~C.~R.~Wijewardhana,
  ``Phase structure of noncompact QED3 and the Abelian Higgs model,''
  hep-ph/0403250.


\bibitem{Gracey:1993sn}
J.~A.~Gracey,
``Electron Mass Anomalous Dimension at O(1/(Nf(2)) in Quantum Electrodynamics,''
Phys.\ Lett.\ B {\bf 317} (1993) 415
[hep-th/9309092].


\bibitem{Gracey:1993iu}
J.~A.~Gracey,
``Computation of Critical Exponent Eta at O(1/\hbox{$N_f$ }$^2$) in Quantum Electrodynamics in Arbitrary Dimensions,''
Nucl.\ Phys.\ B {\bf 414} (1994) 614
[hep-th/9312055].

\bibitem{Braun:2014wja}
  J.~Braun, H.~Gies, L.~Janssen and D.~Roscher,
  ``Phase structure of many-flavor QED$_3$,''
  Phys.\ Rev.\ D {\bf 90} (2014) 3,  036002
  [arXiv:1404.1362 [hep-ph]].
 

\bibitem{Wilson:1971dc} 
  K.~G.~Wilson and M.~E.~Fisher,
  ``Critical exponents in 3.99 dimensions,''
  Phys.\ Rev.\ Lett.\  {\bf 28}, 240 (1972).



\bibitem{Ponte:2012ru}
  P.~Ponte and S.~S.~Lee,
  ``Emergence of supersymmetry on the surface of three dimensional topological insulators,''
  New J.\ Phys.\  {\bf 16} (2014) 1,  013044
  [arXiv:1206.2340 [cond-mat.str-el]].

\bibitem{Dalidovich:2013qta}
  D.~Dalidovich and S.~S.~Lee,
  ``Perturbative non-Fermi liquids from dimensional regularization,''
  Phys.\ Rev.\ B {\bf 88} (2013) 245106
  [arXiv:1307.3170 [cond-mat.str-el]].

\bibitem{Patel}
A.~Patel, P.~Strack, and S.~Sachdev, 
``Hyperscaling at the spin density wave quantum critical point in two dimensional metals,'' 
  arXiv preprint arXiv:1507.05962 (2015).


\bibitem{DiPietro:2015xx} 
 L.~Di Pietro, and E.~Stamou, 
    in progress. 

\bibitem{Grover:2012sp} 
  T.~Grover,
  ``Entanglement Monotonicity and the Stability of Gauge Theories in Three Spacetime Dimensions,''
  Phys.\ Rev.\ Lett.\  {\bf 112}, no. 15, 151601 (2014)
  [arXiv:1211.1392 [hep-th]].


\bibitem{Hands:2002dv} 
  S.~J.~Hands, J.~B.~Kogut and C.~G.~Strouthos,
  ``Noncompact QED(3) with N(f) greater than or equal to 2,''
  Nucl.\ Phys.\ B {\bf 645}, 321 (2002)
  [hep-lat/0208030].
  
  
\bibitem{Hands:2004bh} 
  S.~J.~Hands, J.~B.~Kogut, L.~Scorzato and C.~G.~Strouthos,
  ``Non-compact QED(3) with N(f) = 1 and N(f) = 4,''
  Phys.\ Rev.\ B {\bf 70}, 104501 (2004)
  [hep-lat/0404013].
  
\bibitem{Strouthos:2008kc} 
  C.~Strouthos and J.~B.~Kogut,
  ``The Phases of Non-Compact QED(3),''
  PoS LAT {\bf 2007}, 278 (2007)
  [arXiv:0804.0300 [hep-lat]].

\bibitem{Raviv:2014xna}
  O.~Raviv, Y.~Shamir and B.~Svetitsky,
  ``Nonperturbative beta function in three-dimensional electrodynamics,''
  Phys.\ Rev.\ D {\bf 90} (2014) 1,  014512
  [arXiv:1405.6916 [hep-lat]].


\bibitem{Aharony:2015xx} 
  A much more extensive discussion will appear in O.~Aharony, Z.~Komargodski,~P.Narayan, and T.~Sharma, 
    in progress. 

\bibitem{Polyakov} 
  A.~M.~Polyakov,
  ``Quark Confinement and Topology of Gauge Groups,''
  Nucl.\ Phys.\ B {\bf 120}, 429 (1977).

\bibitem{Brezin:1974eb}
  E.~Brezin, J.~C.~Le Guillou and J.~Zinn-Justin,
  ``Wilson's theory of critical phenomena and callan-symanzik equations in 4-epsilon dimensions,''
  Phys.\ Rev.\ D {\bf 8} (1973) 434.
  
\bibitem{Brezin:1973jc}
  E.~Brezin, J.~C.~Le Guillou and J.~Zinn-Justin,
  ``Approach to scaling in renormalized perturbation theory,''
  Phys.\ Rev.\ D {\bf 8} (1973) 2418.
 
\bibitem{'tHooft:1972fi}
  G.~'t Hooft and M.~J.~G.~Veltman,
  ``Regularization and Renormalization of Gauge Fields,''
  Nucl.\ Phys.\ B {\bf 44} (1972) 189.

\bibitem{Breitenlohner:1977hr}
  P.~Breitenlohner and D.~Maison,
  ``Dimensional Renormalization and the Action Principle,''
  Commun.\ Math.\ Phys.\  {\bf 52} (1977) 11.
 
\bibitem{Akyeampong:1973vj}
  D.~A.~Akyeampong and R.~Delbourgo,
  ``Anomalies via dimensional regularization,''
  Nuovo Cim.\ A {\bf 19} (1974) 219.
 
 \bibitem{Collins}
 J.~C.~Collins, ``Renormalization,'' Cambridge University Press, 1987.
 

\bibitem{Larin:1993tq}
S.~A.~Larin,
``The Renormalization of the Axial Anomaly in Dimensional Regularization,''
Phys.\ Lett.\ B {\bf 303} (1993) 113
[hep-ph/9302240].

\bibitem{Kodaira:1979pa}
  J.~Kodaira,
  ``QCD Higher Order Effects in Polarized Electroproduction: Flavor Singlet Coefficient Functions,''
  Nucl.\ Phys.\ B {\bf 165} (1980) 129.
 

\bibitem{Beneke:1995qq}
M.~Beneke and V.~A.~Smirnov,
``Ultraviolet Renormalons in Abelian Gauge Theories,''
Nucl.\ Phys.\ B {\bf 472} (1996) 529
[hep-ph/9510437].


\bibitem{Xu(2008)} 
C.~Xu,
``Renormalization group studies on four-fermion interaction instabilities on algebraic spin liquids,''
Phys.\ Rev.\ B {\bf 78} (2008) 054432.

\bibitem{Kaveh:2004qa}
  K.~Kaveh and I.~F.~Herbut,
  ``Chiral symmetry breaking in QED(3) in presence of irrelevant interactions: A Renormalization group study,''
  Phys.\ Rev.\ B {\bf 71} (2005) 184519
  [cond-mat/0411594].


\bibitem{Gepner:1984au}
  D.~Gepner,
  ``Nonabelian Bosonization and Multiflavor {QED} and {QCD} in Two-dimensions,''
  Nucl.\ Phys.\ B {\bf 252} (1985) 481.
  
  
\bibitem{Affleck:1985wa}
  I.~Affleck,
  ``On the Realization of Chiral Symmetry in (1+1)-dimensions,''
  Nucl.\ Phys.\ B {\bf 265} (1986) 448.
  
\bibitem{Witten:1983ar}
  E.~Witten,
  ``Nonabelian Bosonization in Two-Dimensions,''
  Commun.\ Math.\ Phys.\  {\bf 92} (1984) 455.
    
\bibitem{Polyakov:2005ss}
  A.~M.~Polyakov,
  ``Supermagnets and sigma models,''
  hep-th/0512310.

 
\bibitem{Gracey:2000am}
  J.~A.~Gracey,
  ``Three loop MS-bar tensor current anomalous dimension in QCD,''
  Phys.\ Lett.\ B {\bf 488} (2000) 175
  [hep-ph/0007171].
   
\bibitem{Dugan:1990df}
M.~J.~Dugan and B.~Grinstein,
``On the Vanishing of Evanescent Operators,''
Phys.\ Lett.\ B {\bf 256} (1991) 239.

\bibitem{Bondi:1989nq} 
  A.~Bondi, G.~Curci, G.~Paffuti and P.~Rossi,
  ``Metric and Central Charge in the Perturbative Approach to Two-dimensional Fermionic Models,''
  Annals Phys.\  {\bf 199}, 268 (1990).
  
\bibitem{Jack:1984vj}
  I.~Jack and H.~Osborn,
  ``General Background Field Calculations With Fermion Fields,''
  Nucl.\ Phys.\ B {\bf 249} (1985) 472.

\bibitem{Huh:2014eea}
Y.~Huh and P.~Strack,
``Stress Tensor and Current Correlators of Interacting Conformal Field Theories in 2+1 Dimensions: Fermionic Dirac Matter Coupled to U(1) Gauge Field,''
JHEP {\bf 1501} (2015) 147
[arXiv:1410.1902 [cond-mat.str-el]].




\bibitem{Chowdhury:2012km}
D.~Chowdhury, S.~Raju, S.~Sachdev, A.~Singh and P.~Strack,
``Multipoint Correlators of Conformal Field Theories: Implications for Quantum Critical Transport,''
Phys.\ Rev.\ B {\bf 87} (2013) 8, 085138
[arXiv:1210.5247 [cond-mat.str-el]].



\bibitem{Giombi:2014xxa}
S.~Giombi and I.~R.~Klebanov,
``Interpolating Between $a$ and $F$,''
JHEP {\bf 1503} (2015) 117
[arXiv:1409.1937 [hep-th]].


\bibitem{Fei:2015oha}
L.~Fei, S.~Giombi, I.~R.~Klebanov and G.~Tarnopolsky,
``Generalized $F$-Theorem and the $\epsilon$ Expansion,''
arXiv:1507.01960 [hep-th].

\bibitem{Giombi:2015haa}
  S.~Giombi, I.~R.~Klebanov and G.~Tarnopolsky,
  ``Conformal QED$_d$, $F$-Theorem and the $\epsilon$ Expansion,''
  arXiv:1508.06354 [hep-th].


 
\bibitem{Rattazzi:2008pe} 
  R.~Rattazzi, V.~S.~Rychkov, E.~Tonni and A.~Vichi,
  ``Bounding scalar operator dimensions in 4D CFT,''
  JHEP {\bf 0812}, 031 (2008)
  [arXiv:0807.0004 [hep-th]].
 
\end{thebibliography}
\end{document}